\pdfoutput=1
\documentclass[fleqn,usenatbib]{mnras}

\usepackage{newtxtext,newtxmath}

\usepackage[T1]{fontenc}
\usepackage{ae,aecompl}

\usepackage{graphicx}	
\usepackage{amsmath}	
\usepackage{amssymb}	

\title[GW170817 and the local kilonova rate]{GW170817: implications for the local kilonova rate and for surveys from ground-based facilities}

\author[M. Della Valle et al.]{
Della Valle, M.$^{1,2,3}$\thanks{E-mail: massimo.dellavalle@inaf.it}
Guetta, D.$^{4}$, 
Cappellaro, E.$^{5}$, 
Amati, L.$^{6}$,\newauthor
~Botticella. M.T.,$^{1}$ 
Branchesi, M.$^{7}$, 
Brocato, E.$^{8}$, 
Izzo. L.$^{2}$,\newauthor
~Perez-Torres. M.A.$^{2,9}$,
Stratta, G.$^{10}$, 
\\
$^{1}$Capodimonte Astronomical Observatory, INAF-Napoli, Salita Moiariello 16, 80131-Napoli, Italy\\
$^{2}$Instituto de Astrof\'isica de Andaluc\'ia (IAA-CSIC), Glorieta de la Astronom\'ia, s/n, 18008, Granada, Spain\\
$^{3}$International Center for Relativistic Astrophysics, Piazzale della Repubblica 2, I-65122 Pescara, Italy\\
$^{4}$Department of Physics and Optical Engineering, ORT Braude College, Karmiel 21982, Israel\\
$^{5}$Padova Astronomical Observatory, INAF-Padova, Vicolo dell'Osservatorio 5, 35122-Padova, Italy\\
$^{6}$INAF-IASF, Sezione di Bologna, via Gobetti 101, 40129 Bologna, Italy\\
$^{7}$Gran Sasso Science Institute, Viale F. Crispi 7, L’Aquila, Italy \\
$^{8}$Roma Astronomical Observatory, INAF-Roma, via di Frascati, 33, 00040 Monteporzio Catone, Italy \\
$^{9}$Departamento de Fisica Teorica, Facultad de Ciencias, Universidad de 
Zaragoza, E-50009, Zaragoza, Spain\\
$^{10}$Universit\'a degli studi di Urbino  \textit{Carlo Bo}~Dipartimento di Scienze di Base e  Fondamenti, Via S. Chiara 27, 61029, Urbino, Italy
}

\date{Accepted XXX. Received YYY; in original form ZZZ}

\pubyear{2018}

\hypersetup{draft}
\begin{document}
\label{firstpage}
\pagerange{\pageref{firstpage}--\pageref{lastpage}}
\maketitle
\newcommand{\lsim}{{\, \lower2truept\hbox{
${< \atop\hbox{\raise4truept\hbox{$\sim$}}}$}\,}}
\newcommand{\gsim}{{\, \lower2truept\hbox{
${> \atop\hbox{\raise4truept\hbox{$\sim$}}}$}\,}}
\newcommand{\Ohat}{{{\widehat \Omega}}}
\newcommand{\oneskip}{{\vskip\baselineskip}}

\begin{abstract}
We compute the local rate of events similar to GRB 170817A, which has been recently found to be associated with a kilonova (KN) outburst. Our analysis finds an observed rate of such events of R$_{KN}\sim 352^{+810}_{-281}$ Gpc$^{-3}$yr$^{-1}$. After comparing at their face values this density of sGRB outbursts with the much higher density of Binary Neutron Star (BNS) mergers of 1540$^{+3200}_{-1220}$ Gpc$^{-3}$yr$^{-1}$, estimated by LIGO-Virgo collaboration, one can conclude, admittedly with large uncertainty that either only a minor fraction of BNS mergers produces sGRB/KN events or the sGRBs associated with BNS mergers are beamed and observable under viewing angles as large as $\theta \lsim 40^{\circ}$. Finally we provide preliminary estimates of the number of sGRB/KN events detected by future surveys carried out with present/future ground-based/space facilities, such as LSST, VST, ZTF, SKA and THESEUS.
 
\end{abstract}





\section{Introduction}

In the last decades 
BNS systems have been targets of interest because of their direct link with some of the most relevant topics of modern astrophysics, such as the indirect confirmation of the existence of GWs through radio observations \citep{hulse}, the predicted connection with short Gamma-Ray Bursts (sGRBs) \citep{eichler} and their direct observations from X and $\gamma$ satellites e.g. \citep{gehrels05} and their detection as sources of GWs \citep{Abbott17a,Abbott17b}. During a BNS  merger some sub-relativistic ejecta of mass $\sim 10^{-3}-10^{-2}$ M$_\odot$ is thrown out along the orbital plane at a modest fraction of the speed of light, $\beta=v/c=0.1-0.3$ and rapid neutron capture in the sub-relativistic ejecta e.g. \citep{lattimer,brian10,tanaka13} is hypothesized to produce a KN, an optical and near-infrared signal lasting hours to weeks \citep{li,tanvir13, smartt17,pian17} powered by radioactive decay. Finally, this sub-relativistic ejecta  transfer most of their  kinetic energy ($\sim 10^{50} - 10^{51}$ erg) to the shocked ambient medium. 

The rate of BNS mergers can be constrained in several ways: by modeling the evolution of binary systems through populations synthesis simulations e.g. \citep{sadow08, ben13, ziosi, dominik15,bel16,chrus18}, by the direct measurements of coalescence rates of BNS in the Milky Way from pulsar observations \citep{narayan,kalogera04}, from the cosmic abundance of r-process elements such as Europio \citep{matteucci14,vangioni16} and from gravitational wave observations \citep{Abbott17b}. However, to date, both BNS mergers and sGRBs rates are poorly constrained (see Tab. 1 and Tab. 2). Moreover, it is not yet clear whether all BNS mergers produce sGRBs/KN events or only a fraction of them is able to do it \citep{bel08}. A lower limit to the measurement of the BNS merging rate can be inferred by measuring  the frequency of occurrence of sGRBs, which are believed to occur during the coalescence of two NSs \citep{eichler,berger14,avanzo}. In this paper, we show that the  discovery of the electromagnetic counterpart of GW 170817 (Abbott et al. 2017b and reference therein) can help to significantly constrain this broad range of uncertainty. On the basis of this new rate we provide preliminary estimates of the number of KNe detected in surveys carried out by present and next generation ground based facilities/space missions, such as LSST, VST, ZTF, SKA and Theseus. 

\section{The rate of BNS merging}

The rate of expected merging events has been estimated based on the observed galactic population of double NS binaries containing a radio pulsar \citep{phinney91,narayan,kalogera01,burgay03}. A reference number for the BNS merger rate  in the Galaxy is $\sim 80^{+200}_{-60}$ Myr$^{-1}$, that can be converted to  $800^{+2000}_{-600}$ Gpc$^{-3}$yr$^{-1}$ for a galaxy number density of $10^{-2}$ Mpc$^{-3}$ \citep{kalogera04}. This rate has been recently revised by Chruslinska et al. (2018) (see Table 1). Population synthesis studies of binary systems give results consistent with the above rates \citep{perna02,bel02,bel07}. Taking into account the BNS merger rate coming from the detection of GW 170817 \citep{Abbott17a} and the design sensitivity of Advanced LIGO and Virgo, gravitational signals from BNS mergers are expected to be detected at a rate of one every 4-90 days \citep{Abbott18}.

\begin{table}
  \caption{BNS merging rate (Gpc$^{-3}$ yr$^{-1}$)}
  \label{tab. 1}
  \begin{tabular}{llll}
     central  & confidence & confidence &Ref.\\
      value & interval & level& \\

    \hline
	800  & 140-2800 & 95$\%$& Kalogera et al. 2004\\
	 316& 100-1000& --& Belczynski et al. 2008\\
       866  & 500-1500& --& Petrillo et al. 2013\\
 	1540&  320-4740 & 90$\%$& Abbott et al. 2017b\\
      --&  300-1200&-- & Chruslinska et al. 2018\\
1109 & 269-3981 & 68$\%$ &Jin et al. 2018\\
\hline
  \end{tabular}
\end{table}

\section{The rate of sGRBs}

In this section we review the estimates of the local rate of sGRBs. In spite of many efforts the rate of sGRBs is still uncertain within a factor $\sim 10^3$, as it appears from an inspection of Tab. 2. There are many parameters that affect the estimate of the rate of sGRBs, such as: i) the shape of the luminosity function assumed for the sGRBs; ii) the minimum of the luminosity function; iii) the redshift distribution assumed for the sGRBs; iv) the time delay distribution of the merging time (from the formation of the NS to their merging); v) the beaming factor\footnote{f$_b^{-1}=1-cos\theta$, where $\theta$ is the jet half-opening angle}. Among these the minimum of the sGRB Luminosity function $L_{\rm min}$, is the parameter that mostly affects the local rate, as it can be clearly seen in the equation below that has been derived by \citet{dafne09}

\begin{equation}
R_{sGRB} \sim \left(\frac{R_0}{Gpc^{-3}\,yr^{-1}}\right)\left(\frac{L_{\rm min}}{10^{49}}\right)^{-0.5}
\end{equation}

Where $R_0$ is the central value given in Table 2.
In Equation 1, it is assumed that the luminosity function of sGRBs can be parametrized by a broken power law. The power law index for the low luminous part that best reproduced the data  is $\alpha\sim 0.5$ (Guetta \& Piran 2005). 
Several authors have tried to estimate the local rate of sGRBs by using the observed peak flux distribution and redshift distribution of the sGRBs, but still the uncertainties are large. 

\subsection{Rates smaller than 1 Gpc$^{-3}$ yr$^{-1}$}

In this subsection we describe briefly the method used by different authors 
to derive the local rate of sGRBs. Guetta \& Piran (2005) fitted the properties of a simulated sGRB population, described by the parametric luminosity and redshift distribution, to a set of observational constraints derived from the population of sGRBs detected by BATSE. On the other hand BATSE was insensitive to low luminous GRBs and in view of Eq. 1, this means to increase the value of $L_{\rm min}$ and to decrease the local sGRB rate. Ghirlanda et al. 2016 find a similar result after selecting a "bona fide" sample of 211 Fermi/GBM bursts with a peak flux larger than 5 ph cm$^{-2}$ s$^{-1}$ aimed at  avoiding the incompleteness effects affecting observations close to the detection threshold. By performing this selection Ghirlanda et al. 2016 consider only the bright sGRBs ($\gsim 10^{50}$ erg) which corresponds to increase $L_{\rm min}$ in Eq. 1 and therefore to decrease the local sGRB rate. Both Guetta \& Piran 2005 and Ghirlanda et al. 2016 assume that sGRBs follow the star formation rate (SFR).

\begin{figure}
	\includegraphics[width=\columnwidth]{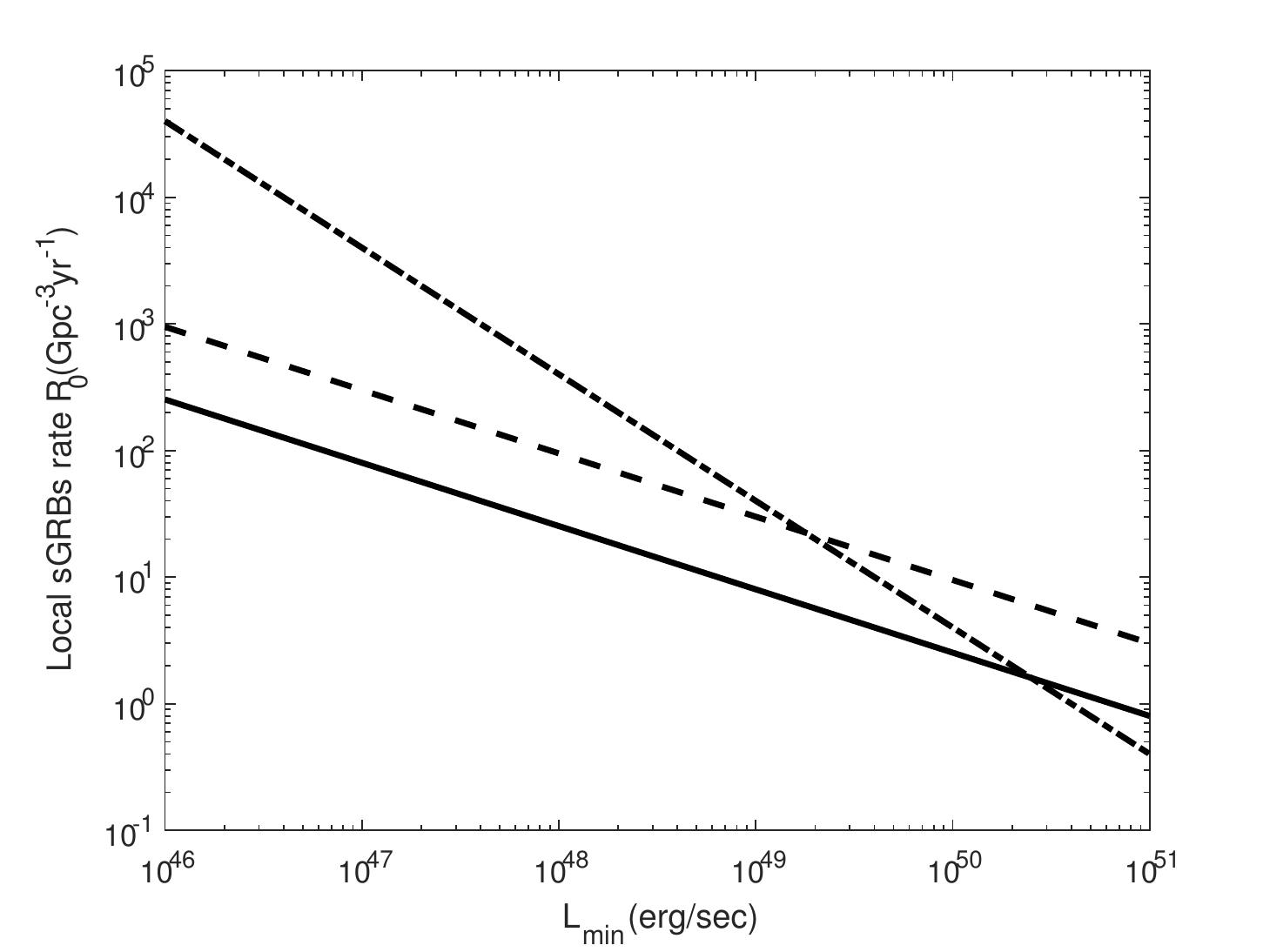}
   \caption{The "observed" local rate of sGRBs as a function of the minimum luminosity as computed by Guetta \& Piran 2006. The rates for an uniform distribution in the delay time (solid line) and strong evolution in the delay time (dashed line) are reported. The dash-dotted line is derived considering Eq. 2 of Nakar et al. 2006.}

\label{fig:example_figure}
\end{figure}

\begin{table}
  \caption{sGRBs rate Gpc$^{-3}$ yr$^{-1}$ per f$_b^{-1}=1$} 
  \label{tab. 2}
  \begin{tabular}{ccc}
     central value & confidence interval & Ref.\\

    \hline
	10   &  5-18    &  Guetta \& Piran (2006) \\
       0.6 &  0.3 - 9 &  Guetta \& Piran (2006) \\
       0.11 &  0.07- 0.18 & Guetta \& Piran (2005) \\
       40   & 10-5$\times 10^5$ & Nakar et al. (2006) \\
       30 &  10-80 &  Guetta \& Piran (2006) \\
       8 &  4-48 &  Guetta \& Piran (2006) \\
	   2 &  1-4 &  Guetta \& Stella (2009)\\  
      8       &   5 -13 &  Coward et al. (2012) \\
     40       &    not given    &  Petrillo et al. (2013)\\
     16.3    &  8.1 - 32.6 & Jin et al. (2015) \\
      11     &   4 - 74 & Fong et al. (2015) \\
      0.8    &   0.65 - 1.1& Ghirlanda et al. (2016) \\
      0.2    &   0.13 - 0.24& Ghirlanda et al. (2016) \\ 
      4.1    &   2.7 - 5.9 & Wanderman \& Piran (2015)\\

\hline
  \end{tabular}
\end{table}

\subsection{Rates larger than 1 Gpc$^{-3}$ yr$^{-1}$}

Several authors find a larger local rate for the sGRBs performing
the same analysis of \citep{dafne05} but considering different sample of sGRBs. Guetta \& Piran 2006 consider a sample of sGRBs detected by Swift. The thereshold of Swift is lower than BATSE allowing the detection of low luminous sGRBs. As it is clear from Eq. 1, the lowering of $L_{\rm min}$ implies an increasing in the local rate. If sGRBs are linked to BNS mergers (see e.g. \citep{narayan01}, the sGRB rate should be given by the convolution of the star formation rate with the distribution Pm($\tau$) of the merging time $\tau$ of the binary system \citep{piran92,ando04}. Considering different delay time distributions and assuming $L_{\rm min} \sim 10^{49}$ erg. Guetta \& Piran (2006) find a range of local rate of $R_0=8-30$ Gpc$^{-3}$ yr$^{-1}$. The different values reported in Table 2 for Guetta \& Piran 2006 correspond to fifferent sGRB redshift and delay-time distributions considered in the paper. These values can be increased if we decrease the minimum luminosity function assumed, as it is apparent from Fig. 1. The form of the luminosity function also affects the local rate of sGRBs (Nakar et al. 2006). The latter authors use a power law function for the luminosity function instead of a broken power law:

\begin{equation}
R_{sGRB} \sim \left(\frac{R_0}{40 Gpc^{-3}yr^{-1}}\right)f_b^{-1}\left(\frac{L_{\rm min}}{10^{49}}\right)^{-1}
\end{equation}
\medskip

and this implies a higher local rate and stronger dependence on the lower value of the luminosity function with respect to Eq. 1. 
Other authors, like \citep{coward12} avoid using an sGRB luminosity function, models for progenitor
rate evolution and a beaming angle distribution. Instead, they focus on observed and measured
parameters that take into account selection effects that modify
the Swift detection sensitivity to sGRBs.
When considering the whole Swift sGRB sample  Wanderman \& Piran 2015 find that it is composed of two populations: one group that are delayed relative to the global SFR with a typical delay time of a 3-4 Gyr (depending on the SFR model) and a second group that follow the SFR with no delay. These two populations are in very good agreement with the division of sGRBs to non-Collapsars and Collapsars suggested by Bromberg et al 2013.
They find a local rate for the non-Collapsar sGRBs given by
\begin{equation}
R_{sGRB} \sim \left(\frac{R_0}{4.1 \pm 0.5Gpc^{-3}yr^{-1}}\right)f_b^{-1}\left(\frac{L_{\rm min}}{5\times 10^{49}}\right)^{-0.95}
\end{equation}
 
Guetta \& Stella (2009) consider the possibility that a fraction of sGRBs come from the merging of dynamically formed binaries in globular clusters and infer the corresponding GW event rate that can be detected with Advanced LIGO/Virgo. In this paper the authors show that a substantial fraction of sGRBs may occur at low redshifts, where the merging of systems formed in globular clusters through dynamical interactions is expected.
They find, under this hypothesis, a merging rate larger than what expected if sGRBs come only from the merging of primordial NSs \citep{sadow08, ben13, ziosi, dominik15,bel16,chrus18}, of the order of $\sim 4$ Gpc$^{-3}$ yr$^{-1}$. However it seems that GW170817 came from an isolated binary system \citep{levan}.

 \section{Upper limits to the KN rate from observations}

In the last decade several sky surveys have been designed for the detection of different classes of transients such as Supernovae, Asteroids, Novae etc. Some of them as ATLAS \citep{tonry11}, DES \citep{doctor17} and DLT40 \citep{yang17} have been used to constrain the rate of KN events. In tab. 3 we report the estimates provided by different teams.  

 \section{ The rate of GRB 170817A-like events }

GRB 170817A and GW170817 events were detected by Fermi-GBM (Fermi-GBM
2017) and INTEGRAL, and  by Advanced LIGO and Advanced Virgo
experiment. The two events, the binary coalescence and the sGRB, were detected  $\sim 1.7s$ apart \citep{Abbott17a,Savchenko17}. About 11h later  the optical counterpart was discovered by \citep{coulter17} in the outskirt of the early type  galaxy NGC 4993, located at $\sim 41$ Mpc \citep{cantiello}. Within one hour and before the announcement of discovery, five other team reported independent detection of the 
optical transients \citep{arcavi17,cowperthwaite17,lipunov17, tanvir17,valenti17}.

In the following we will use the detection of a sGRB/KN event within 41 Mpc to constrain  the rate of such sources. The GRB light curve shows a weak short pulse with a duration $T_{90}$ of about 2 s (50-300 keV).
The time-averaged spectrum is well fit by a power law function with an exponential high-energy cutoff.  The power law index is $\alpha=-0.89 \pm 0.5$.
The 1.024-sec peak photon flux  in the 50-300 keV band
is $F\sim 0.74 ph/s/cm^{2}$. This flux can be compared with the GBM 
peak flux threshold of $F_T\sim 0.50 ph/s/cm^{2}$, then deriving the maximum distance ($D_{\rm max} =49 Mpc$) to which the event could be detected, corresponding to the maximum volume ($V_{\rm max} = 4.9\times 10^{-4}  Gpc^{-3}$). After considering the Fermi-GBM (similary to the INTEGRAL/SPI-ACS) a low-orbit all-sky monitor (e.g. Meegan et al. 2009), with a sky coverage of $S_{\rm cov}\sim 0.64$ and the number of years of operation ($T \sim  9 yr$) of the GBM detector, the rate of sGRBs similar to the 170817A is:
\begin{equation}
R_{sGRB}=\frac{1}{V_{\rm max}}\frac{1}{S_{\rm cov}}\frac{1}{T}\sim 352~Gpc^{-3} yr^{-1}.
\end{equation}

The range of uncertainty on this rate is obviously large, however, in view of the figures reported in Tab. 2 and 3, it still provides interesting clues. On the basis of poissonian statistics applied to one positive detection \citep{gehrels86} we derive an "observed" sGRB-KN rate  of 352$^{+810}_{-281}$ Gpc$^{-3}$yr$^{-1}$ that represents a substantial lower rate than the upper limits derived from the surveys reported in Tab. 3.

\begin{table}
  \caption{Upper limits on KN rate}
  \label{tab. 3}
  \begin{tabular}{cccc}
     Survey & Rate Gpc$^{-3}$ yr$^{-1}$ & c.l.&Ref.\\

    \hline
	ATLAS	& $< 3.0 x 10^4$  & 95\% & \citep{smartt17} \\

	DES  	& $< 2.4 x10^4$   & 90\% & \citep{doctor17} \\

	DLT40       & $<9.9x 10^4$  & 90\%  & \citep{yang17}\\
\hline
  \end{tabular}
\end{table}

\section{Discussion }

To convert the ``observed'' rate into the ``true'' rate of events, one has to correct the observed number for the  beaming factor $f_b^{-1}$. Unfortunately the value of $f_b^{-1}$ is not well known, ranging from values as high as $f_b^{-1}=700-180$ corresponding to angles as small as $\theta =3^\circ-6^\circ$ \citep{ghirla16}, to intermediate values of $f_b^{-1}=30-40$ corresponding to moderately wide angles of $\theta \sim 15^\circ$ \citep{petrillo13,fong15} up to small beaming factors $f_b^{-1} \sim 10$ for angle as broad as  $\theta \sim 30^\circ $ \citep{rez11, granot}.  

By comparing at their face values the density of BNS mergers and GRB 170817A-like events, i.e. 1540$^{+3200}_{-1220}$ Gpc$^{-3}$yr$^{-1}$ \citep{Abbott17a} vs. 352 events Gpc$^{-3}$yr$^{-1}$ we find a range of values for the angles from which a sGRB is observable of $\sim 10^\circ - 40^\circ$ in good agreement with most theoretical predictions (e.g. Lazzati et al. 2017). However the geometry of the outflow of GRB 170817A is still matter of discussion. The radio and X-rays that followed GW170817 are unlike any afterglow observed before, showing a gradual rise over $\sim$ 100 days. They resemble the radio flare predicted long ago to follow BNS mergers. This emission arises from the interaction of the merger outflow with the external medium  \citep{nakar11}. Recently, \citet{nakar18} have considered this model to explain the X-radio observations of GW170817 and they also infer that we have observed an on-axis emission of a ``structured'' jet, a conclusion which might appear in mild disagreement from that derived by \citep{troja17}, who support the idea of an off-axis view for GRB 170817A. 
Very recently Mooley et al. 2018 have found that the late-time emission of GRB 170817A is dominated by a narrowly collimated jet, characterized by an opening angle of 
$\theta\sim 5^{\circ}$ and they conclude that the emission from the jet was likely observed from a viewing angle of $\sim 20^{+5}_{-5}$ degrees. 

If we assume the distance of $\sim 200$ Mpc as a plausible threshold for future detection of GWs associated with BNS mergers in the advanced LIGO/Virgo experiments \citep{Abbott18}, we expect to detect within the corresponding volume of $3.4x10^{-2}$ Gpc$^{-3}$, about N$_{KN}=f_b^{-1} \times 12^{+27}_{-10}$ KNe yr$^{-1}$.

\section{Kilonovae in Future LSST, VST, ZTF, SKA and Theseus Surveys} 

The relatively high rate of sGRB ``170817A-like'' that we have measured with this paper, from dozens to hundredths of events per Gpc$^3$ per year, suggests that future optical sky surveys,  designed with an appropriate observational strategy, will be able to detect a significant number of KNe without GRB or GW triggers. In the following we will provide some preliminary estimates for the KNe detections from LSST, VST, ZTF, SKA and THESEUS surveys. 

\subsection{LSST}

We assume a magnitude limit of R$\sim 24$ for a typical $2 \times 15$s visit and $\sim 20,000$ degs square patrolled in one year (LSST White Paper https://www.lsst.org/scientists/scibook) and 
M$_R \sim -16$ \citep{valenti17} for the absolute magnitude at maximum of the KN associated with GRB 170817A. From this data we derive a sampled volume of $\sim 2.1$ Gpc$^3$ yr$^{-1}$ and therefore a number of KN detections of $\eta~\times~ 150-2440$ being $\eta$ the factor that accounts for the "efficiency" of the survey, which depends on several parameters, such as control time (i.e. a quantity that depends on the survey cadence and the photometric time scale evolution of the transient, see Cappellaro et al. 2015), sky conditions, technical downtime, scheduling constraints. After an assumption  of  $\eta \sim 50\%$ we find that the optical monitoring of the sky with LSST can discover $\sim 75-1220$ KN events (GW170817-like) per year within z$\sim 0.25$. 

\subsection{VST}

After assuming a limiting magnitude of the survey of R$\sim 22.5$ in $t_{exp} \sim 1$ min \citep{brocato, grado} and M$_R \sim -16$ \citep{valenti17} for the absolute magnitude at maximum of a 170817-like KN, and a patrolled field of 10,000 square degrees we find a sampled volume of 0.13 Gpc$^{-3}$, which implies, for $\eta = 0.50$, a number of "observable" KNe of $\sim 5$ - $\sim 76$ KNe/yr. However, an observational strategy based on a survey devoted to the continuous patrolling of known galaxy clusters within hundreds of Mpc would increase the number of KN detections. For example, with a field of view of $\sim 1$ square degree, VST can monitor a large sample of galaxies ($\geq 70$) of the Hydra cluster with only 4 pointings. Considering a slightly different point of view, VST is expected to play a key role during the next O3 observing run of LIGO/Virgo collaboration (LVC). In fact, the O3 network of three interferometers is expected to reduce the uncertainties of the localization of GW events to few tens of square degrees with respect with previous runs. This allows VST to cover $\geq 90\%$ of the LVC skymap probability largely increasing the efficiency in discovering KN candidates and in providing observational constraint on the rate of nearby ($\leq200$ Mpc) KN events.

\subsection{ZTF}

The Zwicky Transient Facility (ZTF) is a new-generation time-domain survey currently operating at the Palomar Observatory. We assume a magnitude limit of R$\sim 20.5$ for a typical exposure of 30s \citep{bellm2017} and $\sim 20,000$ square degrees of sky continuously patrolled \footnote{http://www.ztf.caltech.edu/page/technical} and M$_R \sim -16$ for the absolute magnitude at maximum of a KN \citep{valenti17}. We derive a patrolled volume of $\sim 1.6 \times 10^{-2}$ Gpc$^3$, and therefore a number of kilonova detections of N$_{KN}= \eta~\times~ 5.6^{+13}_{-4.5}$ KNe yr$^{-1}$. 

\subsection{SKA}

The Square Kilometre Array (SKA) will survey large areas of the sky with close to $\mu$Jy sensitivities, which in principle would allow also to detect KNe from the already programmed SKA surveys.  For the sake of simplicity, we discuss here three continuum generic surveys planned with the SKA1 (see Table 4). For each survey, we assume that the sampled area and the rms are obtained after one year of observations,  and 3$\times$10 min visits of the same field to reach the quoted sensitivity. We assume that the peak of brightness of GW170817 at 2.5 GHz  was $S_\nu \approx 100 \mu$Jy e.g.\citep{margutti18}, corresponding to a monochromatic radio luminosity of $L_\nu \approx 2.0\times 10^{26}$ erg/s/Hz at the distance of its host galaxy.  GW170817 peaked in the radio (at $\sim$ 2.5 GHz) around day 163.  Given the relatively standard radio synchrotron behaviour of the KN in GW170817, the peak at the nominal frequency of SKA  of 1.7 GHz is expected to occur around day $\sim 163 \times (2.5/1.7) \approx 240$ d. Therefore, three visits over a year are enough to reliably detect any new KN in the radio. Hence, the factor $\eta$ that accounts for the efficiency of the survey can be safely assumed to be very close to one. However, the sensitivity of the radio surveys forces us to restrict the search for KNe only inside the ''local'' universe. After assuming a 5-$\sigma$ threshold for a bona fide detection in a blind survey, the maximum distance to which we can confidently detect new objects is of $\approx$82, 193, and 373 Mpc  for SKA1-Mid-A, SKA1-Mid-B, and SKA1-Mid-C, respectively. The volumes sampled by the planned SKA1 surveys  after one year are just a tiny fraction of a  Gpc$^3$ (see Table 4). 
For a KN rate of 352$^{+810}_{-281}$ Gpc$^{-3}$\,yr$^{-1}$ (see Section 5) the expected rate of kilonovae to be detected by blind radio surveys such as the ones described above is therefore quite meager. The values reported in col. 5 of Tab. 4 are in fact very similar, since the number of detected KNe scales as follows (P\'erez-Torres et al. 2015): $N_{\rm det} \propto D_{\rm
mas}^3/\Omega \propto \sigma_\nu^{-3/2}/\Omega$, where $\sigma_\nu$ is the sensitivity of the survey and $\Omega$ the angular area covered. Plugging in the numbers for each of those surveys, we see that only the SKA1-Mid-A offers some chances of getting at most one (blind) detection of a KN after one year of observations. SKA1-Mid-B and SKA1-Mid-C, despite being significantly more sensitive than SKA1-Mid-A, cannot compensate the small area covered by those surveys.

\begin{table*}
  \label{tab:SKA}
\begin{tabular}{lrrrr}
\hline 
SKA survey &	Area      & rms            & Sampled Volume & $N_{\rm det}$  \\
&    (sq. deg) & ($\mu$Jy/beam) &  Gpc$^3$ yr$^{-1}$ & (Gpc$^{-3}$\,yr$^{-1}$)   \\
\hline
SKA1-Mid-A &  31,000 &  5.0   & 17$\times10^{-4}$ &  $0.61^{+1.39}_{-0.36}$\ \\	
SKA1-Mid-B &     500 &  0.9   & 3.6$\times10^{-4}$ &  $0.13^{+0.29}_{-0.08}$\ \\	 	
SKA1-Mid-C &      20 &  0.24  & 1.1$\times10^{-4}$ &  $0.04^{+0.08}_{-0.02}$\ \\	 	
\hline
\end{tabular}
\caption{\small \em Expected blind detections of KNe from planned SKA Generic Continuum Extragalactic Surveys}
\end{table*}

\subsection{THESEUS}

The Transient High-Energy Sky and Early Surveyor (THESEUS) space mission \citep{Amati18}, under study by ESA for a possible launch in 2030--2032, would be a perfect GRB and transients machine, capable of providing detection, accurate localization, redshift and characterization of any class of GRBs and, more in general, most classes of transient sources. In particular, THESEUS will provide three observational channels for the e.m. counterparts of BNS and NS-BH mergers: the associated sGRB, the possible soft X-ray emission produced after the merging, the KN emission \citep{Stratta18}. 
The on--board infra-red telescope (IRT) will have a limiting magnitude of $m_H=20.6$ for 300s of exposure. When compared with the absolute magnitude at maximum of a 170817-like KN $M_H=-15.5$ \citep{Stratta18}, we find a maximum detection distance of $\sim0.2$ Gpc.  Based on these expected performances and the rate per co--moving volume and year derived above, we estimate that THESEUS/IRT can detect in one year up to $\sim 2-40$ sGRB-KN events 170817-like. The IRT will perform automatic follow--up observations of the gamma--ray and X--ray sources detected and localized by the Soft X-ray Imager (SXI, 1 sr FOV) and X--Gamma--Ray Imaging Spectrometer (XGIS, FOV $>$2sr). Thus the IRT detection rate of 170817-like KNe quoted above, should take into account the SXI and XGIS detection rate of the high-energy counterpart up to 0.2 Gpc of the same events.  
The luminosities and time--scales of the predicted X-ray emission from   BNS mergers are still very uncertain and for GW170817 no constraints could be obtained since no X-ray monitor was observing during the first minutes-hours. Nevertheless, almost all the proposed models predict fluxes that can be detected by the SXI up to $\sim$0.2 Gpc \citep{Stratta18}. 
The XGIS will detect and localize 30--40 on-axis sGRBs yr $^{-1}$ up to $z\sim 1$ while off-axis sGRB as the 170817 event will be detected up to $\sim70$ Mpc \citep{Stratta18}. Thus the IRT follow-up of sGRB-KN events 170817-like will be mainly driven by the SXI detection rate. In summary, THESEUS will detect in principle 2-40 sGRB-KN events 170817-like per years both in X-rays with SXI and optical/NIR with IRT.

\section{Conclusions} 

The discovery of the KN associated with GRB 170817A/GW170817 has provided the community with a number of "foods for thought" and some of them have been examined in this paper.
\medskip

i) we find an "observed" rate of sGRBs associated with KNe of 
352$^{+810}_{-281}$ Gpc$^{-3}$yr$^{-1}$, which is definitely larger, by 1-2 orders of magnitude, than most of the estimates reported in Tab. 2 (all rates have been normalized to f$_b^{-1}$ =1). This rate is in relatively good agreement with the value of 190$^{+440}_{-160}$ Gpc$^{-3}$yr$^{-1}$ found by \citep{zhang18}. The difference between the two estimates is mainly due do different D${max}$  adopted in the papers:
49Mpc vs. 65Mpc \citep{zhang18}. From a theoretical point of view these high rates can be reproduced for an appropriate choice of the minimum luminosity of the luminosity function, i.e. $\sim 10^{47}$ erg rather than $10^{50-51}$ erg (see eq. 1 and fig. 1). As an alternative one should consider the possibility that GRB 170817A might be the prototype of a class of peculiar/subluminous sGRBs, associated with KNe, which not necessarily form the main bulk of the sGRBs population. For example the subluminous SN 1987A that was discovered in the "local" universe belongs to a relatively tiny class, $\lsim~10\%$ \citep{pasto} of intrinsically rare subluminous type II SNe.  

ii) the comparison between the value of "observed" KN rate presented in this paper with the much higher density of BNS mergers estimated by the LIGO/Virgo collaboration \citep{Abbott17a} points out that either a minor fraction of BNS mergers produce sGRBs characterized by an isotropic emission or most BNS mergers produce beamed sGRBs that can be observed under viewing angles as large as $\theta \sim 40^\circ$. This second scenario is supported by Mooley et al. 2018 who find for GRB 170817A a $\theta \sim 5^{\circ}$ collimated jet that was likely observed from a viewing angle of $\sim 20^{+5}_{-5}$ degrees.   

iii) An "observed" rate of $\sim 352$ events per Gpc$^{-3}$ yr$^{-1}$ offers excellent perspectives for the identification of the electromagnetic counterpart of BNS events detected during next LIGO/Virgo observing runs. On the basis of our analysis we estimate that future optical surveys will be able to detect, within 200 Mpc,  N$_{KN}=f_b^{-1} \times 12^{+27}_{-10}$ KNe yr$^{-1}$.

iv) The relatively high frequency of occurrence of KNe will allow to some of the future facilities the direct detection of KNe without GW triggers. Particularly LSST will have the capability to discover dozens to hundreds of KNe up to z$\sim 0.25$, that is well beyond the current advanced LIGO/Virgo capabilities. An opposite trend is shown by SKA. Preliminary estimates indicate that this array of radio telescopes will have the capability to carry out wonderful follow-ups for KNe discovered by other facilities, while it will be not so efficient in discovering KNe during "blind" and "shallow" radio surveys.

\section*{Acknowledgements}

MDV would like to thank the Instituto de Astrofisica de Andaluc\'ia for its hospitality and creative atmosphere. The authors thanks the referee for her/his comments that have
helped to improve the presentation of the data. We are also indebted with Pia Astone for her critical reading of the manuscript.




\bibliographystyle{mnras}





 

 

%
%
%
 
 
 
 
\end{document}